\documentclass[%
 reprint,
 amsmath,amssymb,
 aps,
]{revtex4-1}

\usepackage{float}
\usepackage{amsmath}
\usepackage{physics}
\usepackage{changes}
\usepackage{graphicx}
\usepackage{dcolumn}
\usepackage{bm}

\begin{document}

\preprint{APS/123-QED}

\title{Singlet, Triplet and Pair Density Wave Superconductivity in the Doped Triangular-Lattice Moir\'e System}

\author{Feng Chen}
\author{D. N. Sheng}
\affiliation{%
Department of Physics and Astronomy, California State University, Northridge, California 91330, USA}%
\date{\today}

\begin{abstract}
    Recent experimental progress has established the twisted bilayer transition metal dichalcogenide (TMD) as a highly tunable platform for studying many-body physics. Particularly, the homobilayer TMDs under displacement field are believed to be described by a generalized triangular-lattice Hubbard model with a spin-dependent hopping phase $\theta$. To explore the effects of $\theta$ on the system, we perform density matrix renormalization group calculations for the relevant triangular lattice t-J model. By changing $\theta$ at small hole doping, we obtain a region of quasi-long-range superconducting order coexisting with charge and spin density wave
    within $0<\theta<\pi/3$. The superconductivity is composed of a dominant spin singlet $d$-wave and a subdominant triplet $p$-wave pairing. Intriguingly, the $S_z=\pm 1$ triplet pairing components feature pair density waves. In addition, we find a region of triplet superconductivity coexisting with charge density wave and ferromagnetism within $\pi/3<\theta<2\pi/3$, which is related to the former phase at smaller $\theta$ by a combined operation of spin-flip and gauge transformation. Our findings provide insights and directions for experimental search for exotic superconductivity in twisted TMD systems.
\end{abstract}
\maketitle
\textit{Introduction.}---Moir\'e bilayer systems have attracted great attention over the last few years due to their high tunability and capacity to host a wealth of exotic states of matter~\cite{andrei_marvels_2021,balents_superconductivity_2020,kennes_moire_2021}. Since the discovery of superconductivity (SC) and Mott insulating phase in magic-angle twisted bilayer graphene (TBG)~\cite{cao_correlated_2018,cao_unconventional_2018}, other Moir\'e systems have been realized and are under active studies~\cite{liu_tunable_2020,chen_signatures_2019}, including twisted bilayer transition metal dichalcogenides (TMDs)~\cite{zhang_flat_2020,shabani_deep_2021,weston_atomic_2020,devakul_magic_2021,an_interaction_2020,naik_ultraflatbands_2018,zhang_su4_2021,regan_mott_2020,schrade_spin-valley_2019,zhang_moire_2020}. Compared to TBG, twisted bilayer TMDs have the advantages of accommodating flat Moir\'e bands over a much wider range of twist angles and fewer low-energy degrees of freedom, allowing for a simpler lattice model description~\cite{wu_hubbard_2018,wu_topological_2019,pan_band_2020}. Strong correlation effects such as correlated insulating phase~\cite{wang_correlated_2020}, metal-insulator transition~\cite{li_continuous_2021,ghiotto_quantum_2021}, stripe phase~\cite{jin_stripe_2021} and quantum anomalous Hall effect~\cite{li_quantum_2021} have recently been observed in these systems.

Twisted TMD bilayers can be classified into hetero- and homo-bilayers according to whether the two layers are made of the same or different materials. The low-energy electronic degrees of freedom in the former are believed to be described by a generalized triangular-lattice Hubbard model with pseudo-spin SU(2) rotation symmetry~\cite{wu_hubbard_2018,tang_simulation_2020}, whereas in the latter the spin SU(2) symmetry is broken into U(1) by a vertical displacement field due to spin-valley locking and inversion symmetry breaking, and consequently the electron hopping acquires a spin-dependent phase $\theta$~\cite{wu_topological_2019,pan_band_2020,schrade_spin-valley_2019,wang_staggered_2023}.  Note that the standard Hubbard and t-J models on triangular lattices, i.e. $\theta=0$, have exhibited a rich phenomenology enhanced by further-neighbor couplings due to the complex interplay between geometric frustration, quantum fluctuations and hole dynamics~\cite{raghu_superconductivity_2010,jiang_superconductivity_2021,zhu_superconductivity_2023,huang_quantum_2023,wang_doped_2004,baskaran_electronic_2003,motrunich_possible_2004,kumar_superconductivity_2003,chen_unconventional_2013,venderley_density_2019,gannot_hubbard_2020,peng_gapless_2021}. The hopping phase $\theta$ is shown to be widely tunable by the displacement field and thus may serve as a novel control knob of the many-body ground states of twisted TMD homobilayers. The magnetic and superconducting phases under the variation of both carrier density and $\theta$ of the U(1) Hubbard model and/or its closely related t-J model (for strong Hubbard U limit) at/near half-filling have been explored through mean-field calculations, renormalization group analysis, quantum cluster methods and Gutzwiller approximation~\cite{zang_hartree-fock_2021,zang_dynamical_2022,pan_band_2020,zhou_chiral_2022,wu_pair-density-wave_2023,belanger_superconductivity_2022,zegrodnik_mixed_2023}. {However, these methods generally are not accurate in treating the strong electronic correlations present in the model~\cite{qin_hubbard_2022}. Here we implement density matrix renormalization group (DMRG)~\cite{white_density_1992} to accurately capture the ground states on quasi-1D few-leg cylinders, and thus reveal the different ordering tendencies at play and gain some insights into the physics at the 2D limit~\cite{stoudenmire_studying_2012,arovas_hubbard_2022}.} Particularly, DMRG has been applied onto a three-leg cylindrical Moir\'e Hubbard model but only weak SC correlations were observed~\cite{wietek_tunable_2022}. The effective spin-model derived at strong U and half-filling limit was also considered for exploring quantum spin liquid~\cite{kiese_tmds_2022}.

\begin{figure}[ht]
\includegraphics[width=0.48\textwidth]{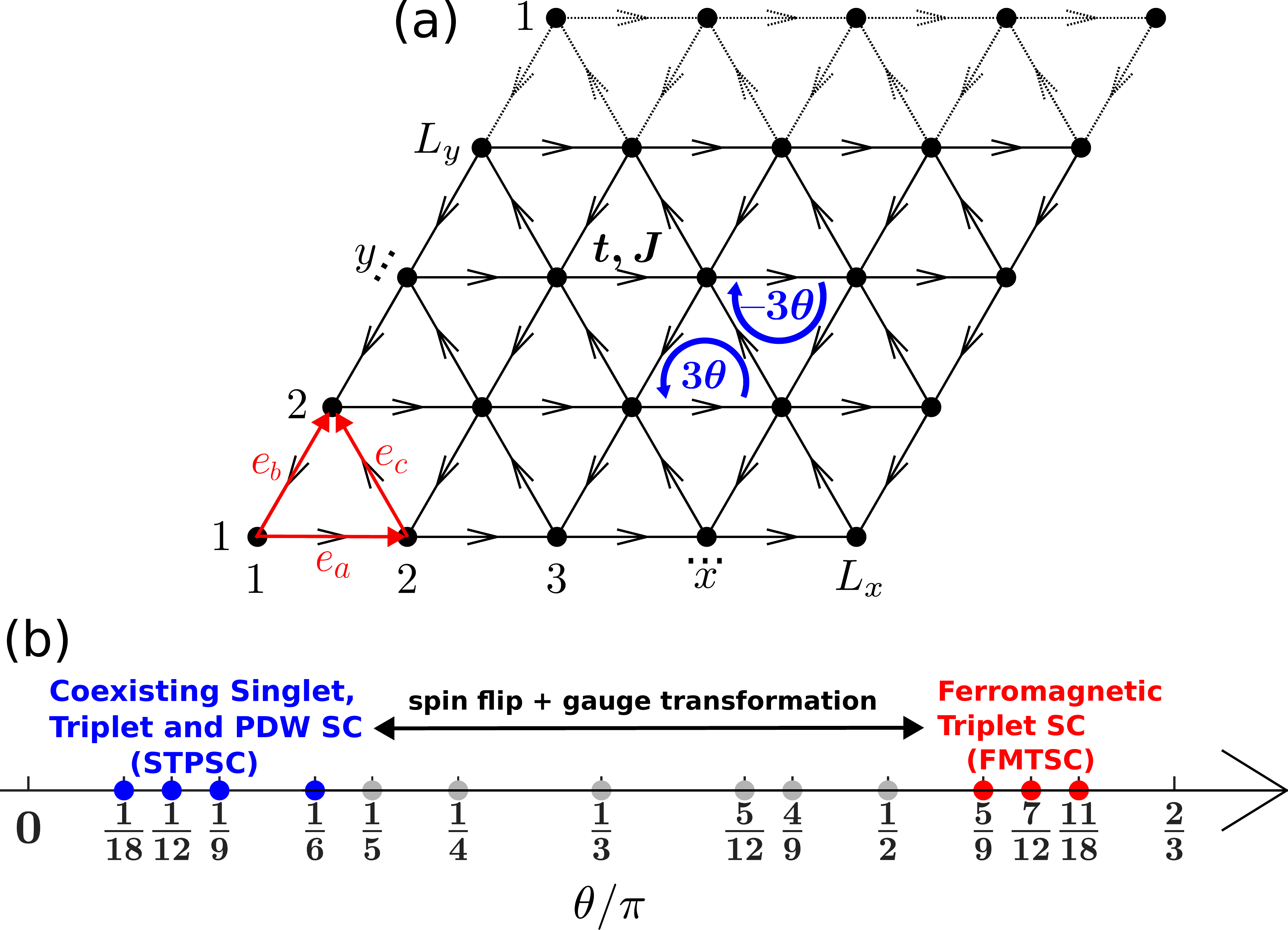}
\centering
\caption{(a) Schematic illustration of the Moir\'e t-J model on a triangular lattice with nearest-neighbor electron hopping ($t$) and spin exchange ($J$). The arrow on each bond is pointed from site $i$ to $j$ in the Hamiltonian Eq.~\ref{Hamiltonian}. It denotes the directional dependence of the hopping phase.
The first and last rows are identified together due to the periodic boundary condition. (b) Quantum phase diagram under the variation $\theta$ for a width-four cylinder. Gray dots denote $\theta$s where no clear signature of SC is observed~\cite{SM}.}
\label{Fig1}
\end{figure}

In this work we study SC of the lightly doped triangular lattice U(1) Moir\'e t-J model on a four-leg cylinder through DMRG calculations.  By varying $\theta$ in the region of $(0,\frac{2\pi}{3})$,  we identify two conjugated superconducting phases as shown in Fig.~\ref{Fig1}(b): (i) Mixed spin singlet $d$-wave and triplet $p$-wave SC coexisting with spin, charge and pair density waves (PDW); (ii) Ferromagnetic triplet $p$-wave SC coexisting with charge density wave (CDW). These two phases are related by a combined operation of spin flip and local gauge transformation, up to a change of the boundary condition.
Their pairing correlations decay algebraically with the Luttinger exponents smaller or around two, demonstrating a robust quasi-long-range SC order~\cite{gong_robust_2021,jiang_high_2021}. Particularly, distinct from other SC phases on the triangular-lattice t-J model~\cite{jiang_superconductivity_2021,huang_quantum_2023,zhu2022}, PDW is a novel SC state where Cooper pairs carry finite center-of-mass momentum~\cite{agterberg_physics_2020},  which are not commonly realized in microscopic models~\cite{berg_pair-density-wave_2010,wu_pair_2023,wu_pair_2023_1,huang_pair-density-wave_2022,jaefari_pair-density-wave_2012,cho_superconductivity_2012,lee_amperean_2014,soto-garrido_pair-density-wave_2014,venderley_evidence_2019,shaffer_triplet_2023}. The plethora of interesting phases found in our calculations could motivate future experimental endeavour in search of novel SC in twisted TMD homobilayers.

\textit{Model and Method.}--- The Moir\'e t-J model is defined as 
\begin{equation}
\begin{split}
  \hat{H}=&-t\sum_{\langle ij\rangle,\sigma=\pm}(e^{-i\sigma\theta}\hat{c}^\dagger_{i\sigma}\hat{c}_{j\sigma}+\text{h.c.})+J\sum_{\langle ij\rangle}(\hat{S}^z_i\hat{S}^z_j\\&+\frac{1}{2}e^{-2i\theta}\hat{S}_i^+\hat{S}_j^-
  +\frac{1}{2}e^{2i\theta}\hat{S}_i^-\hat{S}_j^+-\frac{1}{4}\hat{n}_i\hat{n}_j),
\end{split}
\label{Hamiltonian}
\end{equation}
where $\sigma=\pm$ represents spin up/down, $c^\dagger_{i\sigma}$ and $c_{i\sigma}$ are the creation and annihilation operators for the electron with spin $\sigma$ at the
site $i$, $\langle ij\rangle$ denote nearest neighbors whose locations satisfy $\boldsymbol{r}_j-\boldsymbol{r}_i\in \{\boldsymbol{e}_a, -\boldsymbol{e}_b, \boldsymbol{e}_c\}$ (see Fig.~\ref{Fig1}(a)), $\hat{S}^z_i,\hat{S}^+_i,\hat{S}^-_i$ are the spin-$\frac{1}{2}$ $\hat{z}$ component, raising and lowering operators at site $i$ respectively, and $\hat{n}_i=\sum_\sigma\hat{c}^\dagger_{i\sigma}\hat{c}_{i\sigma}$ is the electron number operator. Double occupancy is prohibited. The hopping phase $\theta$ produces a flux of $\pm 3\theta$ at each triangular plaquette, and a gauge transformation connects two models differing in the fluxes by $2\pi$. 
 We therefore focus on the region of $0<\theta<2\pi/3$. In the present study, we set the hole doping level $\delta=1/12$, and choose $J=1$ and $t=3$, {corresponding to a realistic situation of $U/t=12$~\cite{pan_band_2020}}.

To obtain the ground state, we employ DMRG simulation with U(1)$\times$U(1) symmetry corresponding to charge and spin conservation on a cylindrical system with periodic boundary condition (PBC) along the circumferential ($\boldsymbol{e}_b$ or $y$-) direction and open boundary condition along the axial ($\boldsymbol{e}_a$ or $x$-) direction. The number of lattice sites is given by $N=L_x\times L_y$, where $L_x$ and $L_y$ are the number of sites along $x$- and $y$-direction respectively and are set as $L_y=4$ and $L_x=36$ in the main text. The corresponding geometry is called YC$L_y$~\cite{yan_spin-liquid_2011}. The doping level is defined by $\delta=1-N_e/N$ and we consider the zero total spin-z sector: $\sum_i\hat{S}^z_i=0$, which hosts the ground state as verified in Sec. A of the supplemental materials (SM)~\cite{SM}. In DMRG, the number of Schmidt states kept for representing the
reduced density matrix on either side of the system under bipartition is called ``bond dimension'' $M$~\cite{white_density_1992}.
The calculations improve with the increase of $M$ and become exact for a sufficiently large $M$.

\begin{figure}[htbp]
\includegraphics[width=0.48\textwidth]{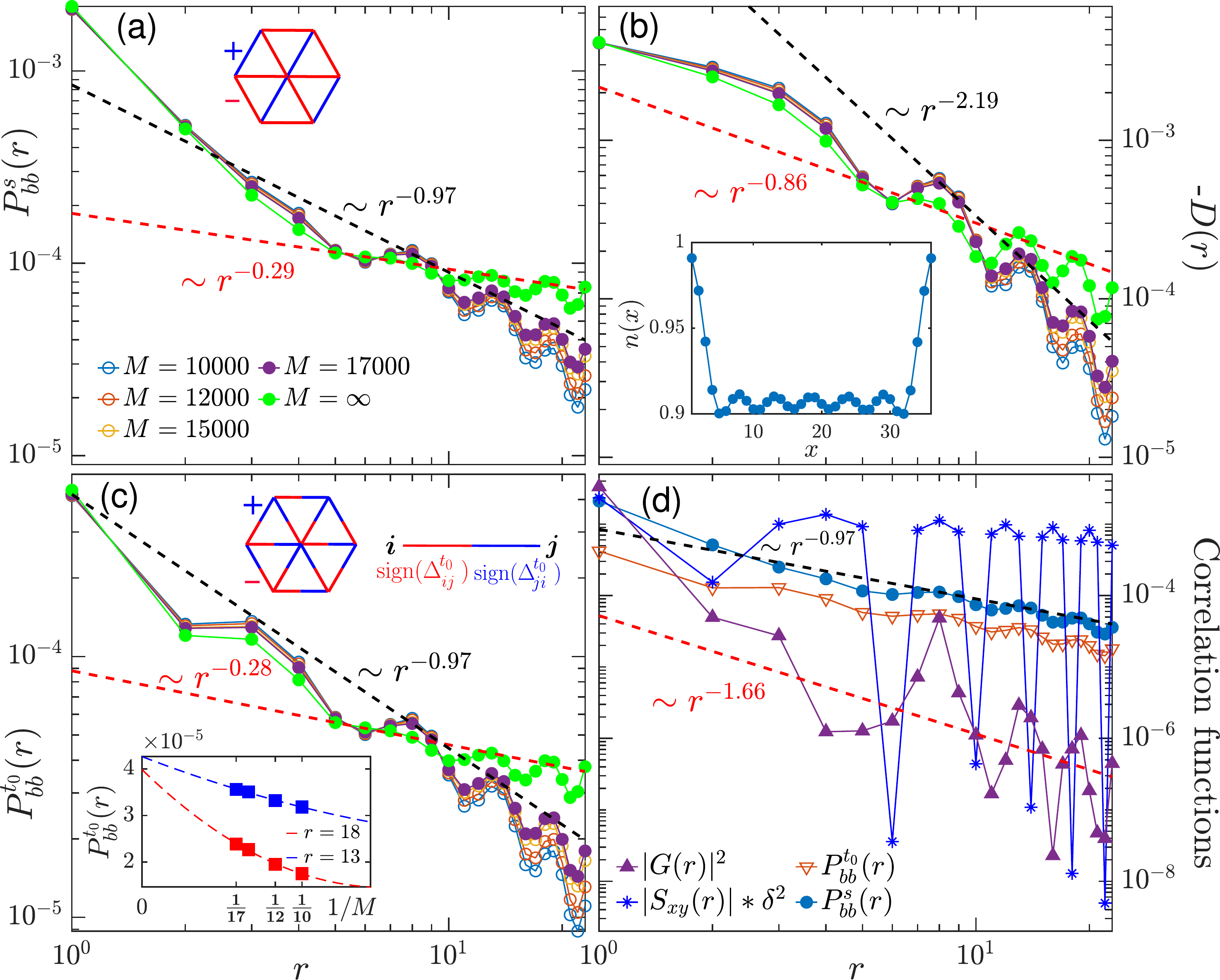}
\centering
\caption{Correlation functions at $\theta=\pi/12$ in the STPSC phase. (a) Scaling of the singlet pairing correlation $P^s_{bb}(r)$ through second-order polynomial extrapolation in terms of inverse bond dimension $1/M$. The extrapolated data at infinite $M$ and $M=17000$ are fitted by power-law decays. The inset shows the relative signs of the pairing order parameters along different bonds, which has a pattern consistent with an ordinary $d$-wave symmetry: {sign($\Delta^s_a$)=sign($\Delta^s_b$)=$-$sign($\Delta^s_c$)}~\cite{SM}. 
(b) The density-density correlation. The inset shows the rung-averaged electron density profile $n(x)=\sum_{y=1}^{L_y}\langle \hat{n}(x,y)\rangle/L_y$ along $\boldsymbol{e}_x$, where charge stripes are observed.  (c) An analogous plot for the triplet paring correlation in the opposite-spin channel $P^{t_0}_{bb}(r)$. The shown sign pattern of the pairing order parameter is consistent with an ordinary $p$-wave symmetry: {sign($\Delta^{t_0}_a$)=sign($\Delta^{t_0}_b$)=sign($\Delta^{t_0}_c$)}~\cite{SM}. Each bond $\langle ij\rangle$ is divided into two halves and the half that includes $i$($j$) is denoted by the sign of $\Delta^{t_0}_{ij}$($\Delta^{t_0}_{ji}$). The sign changes between the two halves because the order parameter is antisymmetric: $\Delta^{t_0}_{ij}=-\Delta^{t_0}_{ji}$. The inset shows an example of data extrapolation to $M=\infty$. (d) Comparison between different correlations at $M=17000$ with the truncation error around $3\times 10^{-6}$. $G(r)$ can also be fitted by an exponential decay with a correlation length around 8.7~\cite{SM}.}
\label{Pi_12_SC}
\end{figure}

\textit{Coexisting Singlet, Triplet and PDW SC (STPSC).}---The SC order is examined by the spin-singlet and triplet pairing correlation functions $P_{\alpha\beta}^s(r)$ and $P_{\alpha\beta}^{t_n}(r)$ defined by
\begin{equation}
    \begin{split}
        P_{\alpha\beta}^s(r)&\equiv\langle \hat{\Delta}^{s,\dagger}_\alpha(\boldsymbol{r}_0)\hat{\Delta}^s_\beta(\boldsymbol{r}_0+r\boldsymbol{e}_x)\rangle\\
        P_{\alpha\beta}^{t_n}(r)&\equiv\langle \hat{\Delta}^{t_n,\dagger}_\alpha(\boldsymbol{r}_0)\hat{\Delta}^{t_n}_\beta(\boldsymbol{r}_0+r\boldsymbol{e}_x)\rangle,
    \end{split}
\end{equation}
where the reference point $\boldsymbol{r}_0\equiv(x_0,y_0)=(L_x/4,L_y)$ and the pairing operators $\hat{\Delta}^{s}_\alpha(\boldsymbol{r}_1)$ and $\hat{\Delta}^{t_n}_\alpha(\boldsymbol{r}_1)$ are defined on the bond along $\boldsymbol{e}_\alpha$ ($\alpha=a,b,c$) at site $\boldsymbol{r}_1$:
\begin{equation}
    \begin{split}
        \hat{\Delta}^{s}_\alpha(\boldsymbol{r}_1)&=(\hat{c}_{\boldsymbol{r}_1\uparrow}\hat{c}_{\boldsymbol{r}_1+\boldsymbol{e}_\alpha\downarrow}-\hat{c}_{\boldsymbol{r}_1\downarrow}\hat{c}_{\boldsymbol{r}_1+\boldsymbol{e}_\alpha\uparrow})/\sqrt{2}\\
          \hat{\Delta}^{t_0}_\alpha(\boldsymbol{r}_1)&=(\hat{c}_{\boldsymbol{r}_1\uparrow}\hat{c}_{\boldsymbol{r}_1+\boldsymbol{e}_\alpha\downarrow}+\hat{c}_{\boldsymbol{r}_1\downarrow}\hat{c}_{\boldsymbol{r}_1+\boldsymbol{e}_\alpha\uparrow})/\sqrt{2}\\
    \hat{\Delta}^{t_{-1}}_\alpha(\boldsymbol{r}_1)&=\hat{c}_{\boldsymbol{r}_1\downarrow}\hat{c}_{\boldsymbol{r}_1+\boldsymbol{e}_\alpha\downarrow},\;\; \hat{\Delta}^{t_{1}}_\alpha(\boldsymbol{r}_1)=\hat{c}_{\boldsymbol{r}_1\uparrow}\hat{c}_{\boldsymbol{r}_1+\boldsymbol{e}_\alpha\uparrow}\;\;.
    \end{split}
\end{equation}
Here $\hat{\Delta}_\alpha^{t_n}$ corresponds to the triplet pairing with total spin-z  $S_z=n$. 

Fig.~\ref{Pi_12_SC}(a) and (c) show two dominant pairing components: $b$-bond singlet pairing $P^s_{bb}(r)$ and opposite-spin-z ($S_z=0$) triplet pairing $P^{t_0}_{bb}(r)$ for $\theta=\pi/12$ in the STPSC phase. Both exhibit power-law decay $P^{s(t_0)}_{bb}(r)\sim r^{-K^{s(t_0)}_{SC}}$ with the Luttinger exponents $K^{s(t_0)}_{SC}\approx 0.3$, suggesting strongly diverging SC susceptibilities $\chi\sim T^{-(2-K_{SC})}$ as the temperature $T\rightarrow 0$~\cite{arrigoni_mechanism_2004}.
Note also that slow power-law decays are already exhibited by the largest-$M$ results with exponents around 0.97. The singlet pairing component is larger in amplitude than the triplet one, and they exhibit $d$-wave and $p$-wave symmetry respectively~\cite{raghu_superconductivity_2010,hsu_topological_2017,venderley_density_2019}. The mixing of singlet and triplet pairings are permitted by the absence of the inversion and spin SU(2) symmetry~\cite{yip_noncentrosymmetric_2014}. In particular, the absence of inversion center allows the mixing of parity-odd $p$-wave and parity-even $d$-wave basis functions in the irreducible representation $E$ of the symmetry group $C_{3v}$ of the system~\cite{hsu_topological_2017}. The charge density correlation function $D(r)\equiv\langle\hat{n}(\boldsymbol{r}_0)\hat{n}(\boldsymbol{r}_0+r\boldsymbol{e}_x)\rangle-\langle\hat{n}(\boldsymbol{r}_0)\rangle\langle\hat{n}(\boldsymbol{r}_0+r\boldsymbol{e}_x)\rangle$ in Fig.~\ref{Pi_12_SC}(b) decays algebraically with a relatively larger exponent (around 0.86), suggesting weaker charge density modulations coexisting with stronger SC. Correspondingly we observe a charge stripe order with two holes per stripe in the inset. For comparison, Fig.~\ref{Pi_12_SC}(d) presents also the in-plane spin-spin correlations $S_{xy}(r)$ defined by
\begin{equation*}
        S_{xy}(r)\equiv \langle \hat{S}^x(\boldsymbol{r}_0)\hat{S}^x(\boldsymbol{r}_0+r\boldsymbol{e}_x)+\hat{S}^y(\boldsymbol{r}_0)\hat{S}^y(\boldsymbol{r}_0+r\boldsymbol{e}_x)\rangle
\end{equation*}
and the Green's function $G(r)\equiv \sum_\sigma\langle \hat{c}^\dagger_{\boldsymbol{r}_0,\sigma}\hat{c}_{\boldsymbol{r}_0+r\boldsymbol{e}_x,\sigma}\rangle$.
The in-plane spin correlation is the strongest among all correlations, characterizing a robust spin density wave order inherited from the 2D in-plane $120^{\circ}$ N\'eel order at half filling based on the spin structure factor calculations~\cite{wu_topological_2019,zang_hartree-fock_2021,SM}. The Green's function squared $\lvert G(r)\rvert^2$ is much weaker than the main pairing correlations, confirming the dominance of two-electron pairing over single-electron tunnelings.

\begin{figure}[ht]
\includegraphics[width=0.3\textwidth]{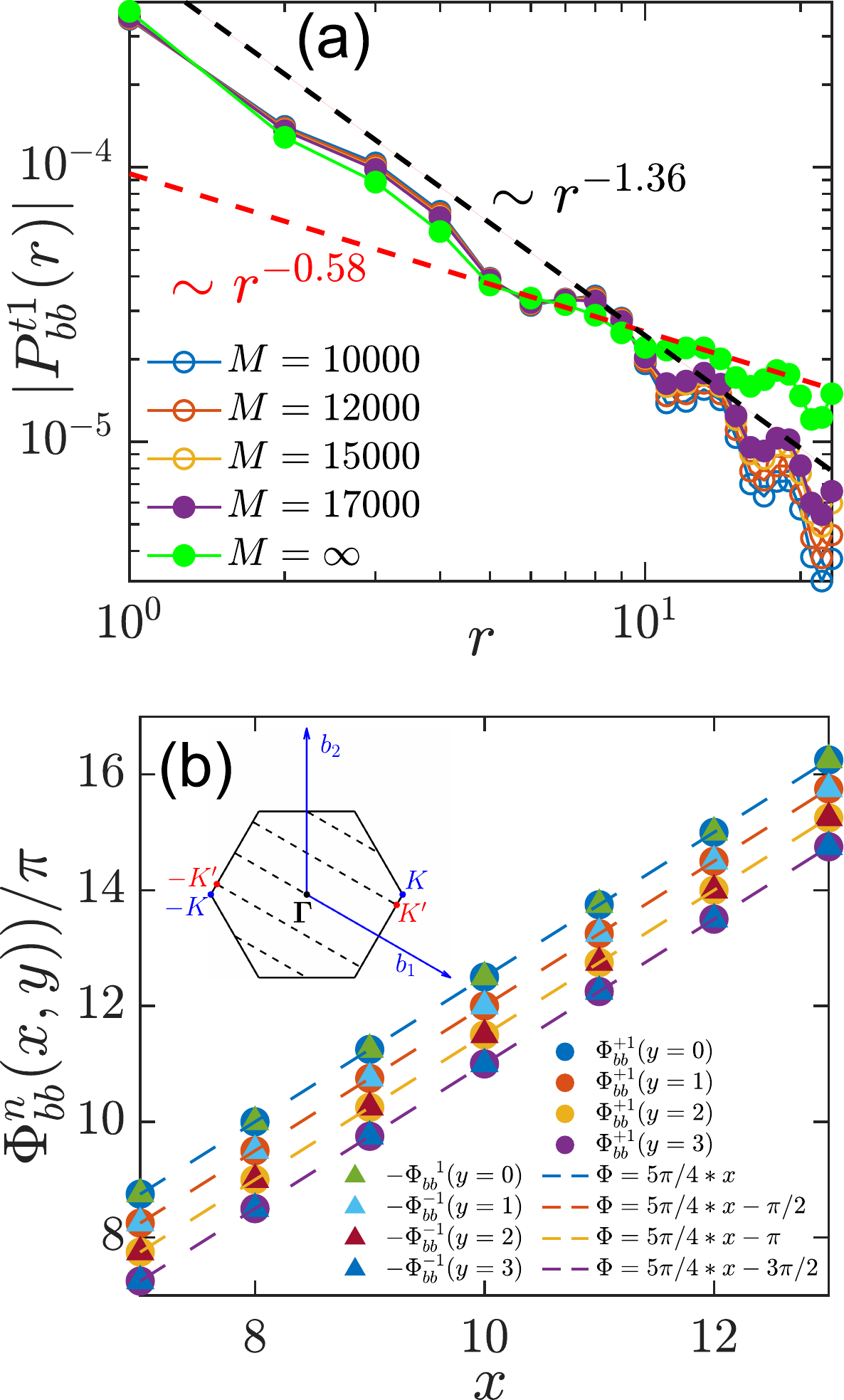}
\centering
\caption{PDW order for $\theta=\pi/12$ in the STPSC phase. (a) Scaling and fitting of the $S_z=1$ component of the triplet pairing correlations. The $S_z=-1$ component is identical due to the time-reversal symmetry; (b) Characterization of spatial phase structure of PDW by $\Phi^n_{bb}(x,y)$.
The wavevectors of PDWs are identified $\boldsymbol{k^{+1}_\text{pdw}}=-\boldsymbol{k^{-1}_\text{pdw}}=\boldsymbol{K'}\equiv\frac{1}{4}\boldsymbol{b_2}+\frac{5}{8}\boldsymbol{b_1}$, where $\boldsymbol{b_{1,2}}$ is the reciprocal wavevector conjugated to $\boldsymbol{e_{a,b}}$. The dashed lines in the inset denote the wavevectors  in the Brillouin zone supported by the YC4 geometry. 
}
\label{PDW}
\end{figure}

Moreover, in the $S_z=\pm 1$ triplet pairing components, we observe quasi-long-range PDW orders with a Luttinger exponent around 0.58 in Fig.~\ref{PDW}(a). The PDW wavevector $\boldsymbol{k_\text{pdw}}$ can be determined by the variation of the phase of the pairing correlation under displacement along both $\boldsymbol{e_a}$ and $\boldsymbol{e_b}$. Specifically,
\begin{equation}
    \begin{split}
        \Phi^n_{bb}(x,y)\equiv&\text{arg}\left(P^{t_n}_{bb}(x\boldsymbol{e_a}+y\boldsymbol{e_b})\right)\\
        =&\text{arg}\left(\langle \hat{\Delta}^{t_n,\dagger}_b(\boldsymbol{r}_0)\hat{\Delta}^{t_n}_b(\boldsymbol{r}_0+x\boldsymbol{e}_a+y\boldsymbol{e}_b)\rangle\right)\\
=&\boldsymbol{k^{n}_\text{pdw}}\cdot(x\boldsymbol{e_a}+y\boldsymbol{e_b})
    \end{split}
\end{equation}
characterizes spatial variation of the phase of the $b$-bond triplet pairing order parameters.
In Fig.~\ref{PDW}(b), $\boldsymbol{k^{{\pm 1}}_\text{pdw}}$ is determined to be $\pm\boldsymbol{K'}$, which are the nearest accessible wavevectors to the Brillouin zone corners $\pm\boldsymbol{K}$ in the YC4 geometry. The same PDW wavevectors are identified for $a$- and $c$-bond. Note that a PDW ground state with $\boldsymbol{k^{\pm 1}_\text{pdw}}=\mp \boldsymbol{K}$ was also predicted for the Moir\'e Hubbard model at $\theta=\pi/3$ by perturbative renormalization group analysis in the weak coupling regime~\cite{wu_pair-density-wave_2023,wu_pair_2023}.

\begin{figure}[!htbp]
\includegraphics[width=0.48\textwidth]{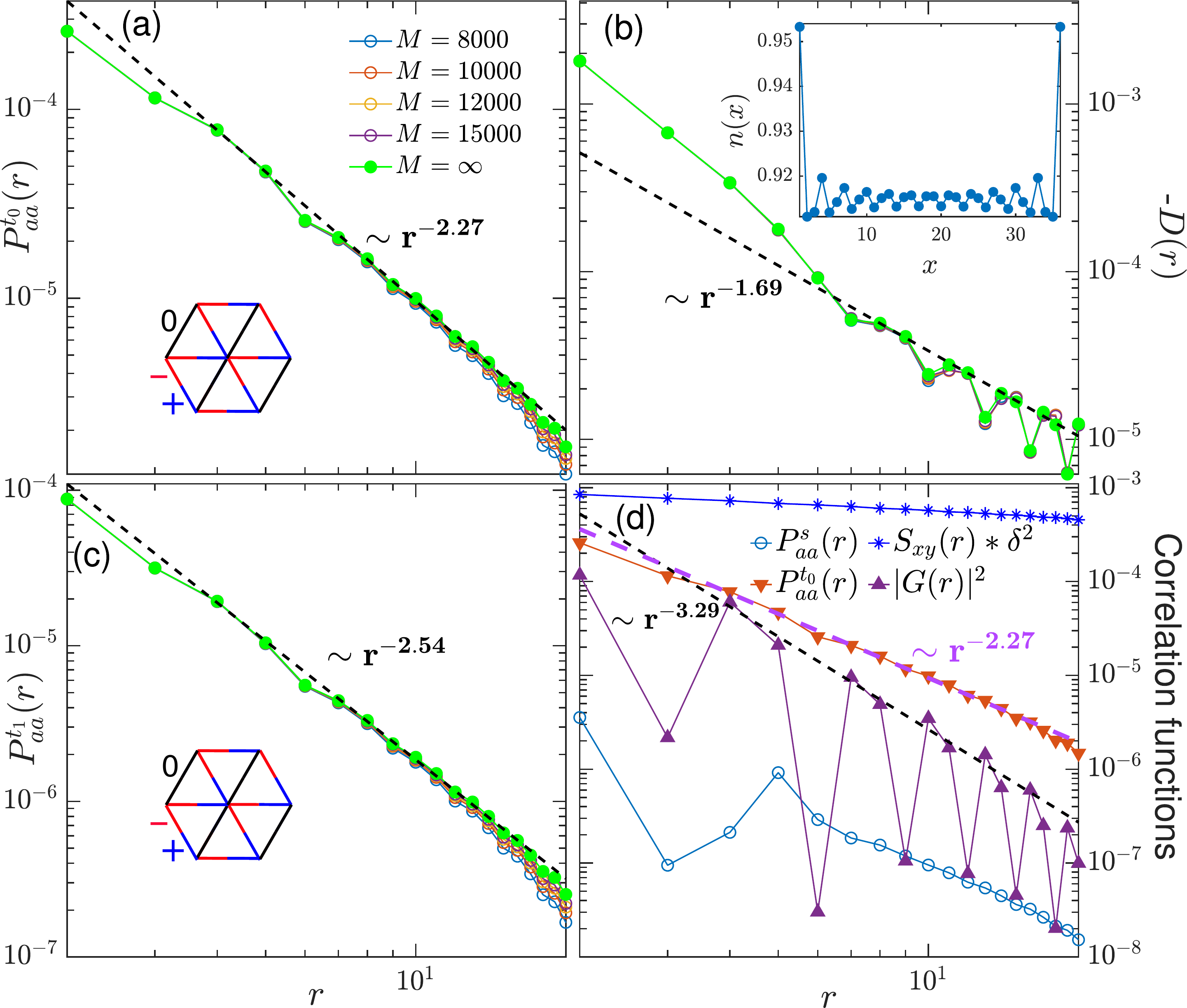}
\centering
\caption{Correlation functions for $\theta=7\pi/12$ in the FMTSC phase, which is conjugated to $\theta=\pi/12$. (a) The spatial decay of $P^{t_0}_{aa}$, which is the strongest opposite-spin triplet pairing correlation among different bonds. The sign structure of the pairing order parameter is consistent with $p$-wave symmetry: {sign($\Delta^{t_0}_a$)=-sign($\Delta^{t_0}_c$), $\Delta^{t_0}_b$=0}~\cite{SM}. The black color denotes a vanishing amplitude. (b) Power-law decay of the charge-density correlation. The inset shows electron density along the axial direction, displaying a charge stripe order.  (c) An analogous plot for the $S_z=1$ component of the triplet pairing correlation $P^{t1}_{aa}$. (d) Comparison between different correlation functions from data at $M=15000$ with the truncation error around $2\times 10^{-7}$. $G(r)$ also fits an exponential decay with a correlation length around 5.4~\cite{SM}.}
\label{FM_TSC}
\end{figure}

\textit{Ferromagnetic Triplet SC (FMTSC).}---In the FMTSC phase, we find the dominant pairing channel to be a $p$-wave spin triplet. In Fig.~\ref{FM_TSC}(a) and (c), both $P^{t_0}_{aa}$ and $P^{t_1}_{aa}$ are non-oscillatory, in accordance with uniform SC order in the bulk of the system, and decay algebraically with exponents slightly larger than 2. An accompanying CDW order is confirmed in Fig.~\ref{FM_TSC}(c) by both the quasi-long-range density correlation ($\sim r^{-1.75}$) and charge stripes in the electron density profile (one hole per stripe). In Fig.~\ref{FM_TSC}(d), a robust in-plane ferromagnetic spin correlation is observed in reminiscence of the parent ferromagnetic order~\cite{wu_topological_2019,zang_hartree-fock_2021}, with the  total spin $S\approx \sqrt{\langle \hat{S}^2\rangle}\approx 0.326N_e$. The singlet paring is shown much weaker than the triplet ones as the triplet pairing is favored by ferromagnetism. {The opposite-spin-z triplet pairing correlation $P^{t_0}_{aa}$ has stronger amplitude and slower decay rate than those of the same-spin-z component $P^{t_1}_{aa}$ because the ferromagnetic order is in-plane.}

\textit{Discussion and Summary}---The FMTSC and STPSC phases are related by a spin-flip operation followed by a local gauge transformation \cite{zhou_chiral_2022} as demonstrated in the SM. Particularly, the uniform z-spin-polarized triplet pairing order at $\theta$ in the FMTSC region is conjugated to the PDW order with $\boldsymbol{k^{\pm 1}_\text{pdw}}=\pm \boldsymbol{K}$ at $(2\pi/3-\theta)$ in the STPSC region:
\begin{equation}
\begin{split}
\Delta^{-\sigma,-\sigma}_\alpha(\frac{2\pi}{3}-\theta,\boldsymbol{r})=e^{-i\sigma\boldsymbol{K}\cdot\boldsymbol{r}-i\sigma\beta_\alpha}\Delta^{\sigma\sigma}_\alpha(\theta),
\end{split}
\label{TF}
\end{equation}
with $\beta_\alpha=\boldsymbol{e_\alpha}\cdot(\boldsymbol{b_1-b_2})/3$, where $\boldsymbol{b_{1,2}}$ are the reciprocal wavevectors conjugated to $\boldsymbol{e_a}$ and $\boldsymbol{e_b}$ respectively. This is consistent with our observations at $\theta=7\pi/12$ (Fig.~\ref{FM_TSC}(c)) and its conjugated partner $\theta=\pi/12$ (Fig.~\ref{PDW}), albeit with a different flux (y-boundary phase) into the 4-leg cylinder. Moreover,
\begin{equation}
\begin{split}
    \Delta^s_\alpha(2\pi/3-\theta)&=-\cos{(\beta_\alpha)}\Delta^s_\alpha(\theta)-i\sin(\beta_\alpha) \Delta^{t_0}_\alpha(\theta)\\
    \Delta^{t_0}_\alpha(2\pi/3-\theta)&=\cos{(\beta_\alpha)}\Delta^{t_0}_\alpha(\theta)+i\sin(\beta_\alpha) \Delta^{s}_\alpha(\theta),
\end{split}
\label{ST_conj}
\end{equation}
 which means that the singlet and opposite-spin triplet pairing components are superposed to produce their counterparts in the conjugated phase. Since the singlet pairing component at $\theta=7\pi/12$ is found negligible compared to the triplet components, and $\beta_\alpha=2\pi/3$ ($\alpha=a,b$) or $4\pi/3$ ($\alpha=c$), one has $\abs{\Delta^s_\alpha(\theta=\pi/12)}\approx\sqrt{3}\abs{\Delta^{t_0}_\alpha(\theta=\pi/12)}$ according to Eq.~\ref{ST_conj}, which explains the larger magnitude of the spin singlet pairing than that of the triplet and the same power-law exponents in Fig.~\ref{Pi_12_SC} (a) and (c).

However, the pairing correlations at $\theta=\pi/12$ has much stronger magnitude (over one order of magnitude larger) and slower decay rate compared to those at $\theta=7\pi/12$ ($K_\text{SC}\approx 0.29$ vs. $K_\text{SC}\approx 2.27$). This in addition to the difference in charge distributions (two vs. one holes per stripe) is caused by the change of the boundary condition: the periodic boundary condition at $\theta=\pi/12$
\begin{equation}
\hat{c}_{\boldsymbol{r}+L_y\boldsymbol{e_y},\sigma}=\hat{c}_{\boldsymbol{r}\sigma}
\end{equation}
turns into a twisted boundary condition~\cite{gannot_how_2023} at $\theta=7\pi/12$
\begin{equation}
\hat{c}_{\boldsymbol{r}+L_y\boldsymbol{e_y},\sigma}=\hat{c}_{\boldsymbol{r}\sigma}e^{i2\pi\sigma L_y/3}
\end{equation}
after the gauge transformation, corresponding to inserting a magnetic flux of $\pm 2\pi L_y/3$ through the interior of the cylinder for electrons. The spin structure factor of the $120^{\circ}$ N\'eel order for $0<\theta<\pi/3$ is peaked at $\pm\boldsymbol{K}$, which are not resolved in the 4-leg cylinder under PBC, whereas for $\pi/3<\theta<2\pi/3$ the system is ferromagnetic with the peak at the system-supported momentum $\boldsymbol{\Gamma}$. Therefore, the former regime is more frustrated than the latter in the YC4 geometry and this might result in stronger SC. The sensitivity of SC to boundary conditions reveal finite-size effects in our four-leg system, {so we also study a different cylinder geometry XC4~\cite{szasz_chiral_2020} (Sec.~H in SM) as well as a YC3 system with $N=40\times 3$ (Sec.~F in SM). Both systems preserve the PBC under local gauge transformation and support $\boldsymbol{\Gamma}$ and $\boldsymbol{\pm K}$ in the Brillouin zone, therefore introducing no frustration.} In the XC4 geometry, we again obtain the STPSC and FMTSC phases and their SC correlations now have similar amplitudes and decay with close exponents ($\approx 2$), consistent with Eq.~\ref{ST_conj}. {In the YC3 cylinder at $\theta=7\pi/12$, the Luttinger exponents for SC ($K^{t_0}_\text{sc}\approx 2.28$) is nearly identical to that of the YC4 cylinder ($K^{t_0}_\text{sc}\approx 2.27$).} The observation of quasi-long-range SC order at different boundary conditions, cylinder geometries and sizes is positive evidence for the existence of SC in the 2D limit~\cite{szasz_chiral_2020}. 

{In contrast with the topological SC phases reported in the mean field and perturbative renormalization group studies of the doped TMD homobilayer~\cite{zhou_chiral_2022,wu_pair-density-wave_2023} or monolayer~\cite{hsu_topological_2017,yuan_possible_2014}, both the $d$- and $p$-wave SC phases found here are topologically trivial as the nearest-neighbor pairings acquire a phase of either 0 or $\pi$ after a $\pi/3$ rotation, instead of the nontrivial phases of $\pm\pi/3$ and $\pm 2\pi/3$ for $p\pm ip$ and $d\pm id$-wave topological SC phases~\cite{huang_topological_2022,jiang_topological_2020,huang_quantum_2023}. Furthermore, the SC phase here is distinct from the Ising SC found in electron-doped TMD monolayers~\cite{zhou_ising_2016,lu_evidence_2015,xi_ising_2016,saito_superconductivity_2016} in that the former arises from hole doping the parent in-plane magnetic Mott insulator at strong electronic couplings whereas the latter the pinning of the electron spins in the Cooper pairs to the out-of-plane directions by the Ising spin-orbit interaction at weak electronic couplings.}  Finally, the $\theta=\pi/6$ case was also studied in Ref.~\cite{wietek_tunable_2022}, but a rather large power-law decay exponent ($\approx 3.34$) was found, so only weak SC was claimed there. Consistently we find that $\theta=\pi/6$ is located at the boundary of the SC  region in Fig.~\ref{Fig1}, and its conjugated pair $\theta=\pi/2$ exhibits no clear signature of SC possibly because of less frustration.

In summary, we perform large-scale DMRG simulations of the Moir\'e t-J model on four-leg cylinders at small hole doping. By varying the spin-dependent hopping phase induced by the out-of-plane electric field, we identify two conjugated SC phases, one of which is characterized by the coexistence of singlet $d$-wave, triplet $p$-wave SC and PDW, and the other ferromagnetic triplet SC. Our study supports twisted TMDs as a highly tunable platform for realizing exotic SC phases.

\textit{Data Availability.---}
The ITensor DMRG code and the data for all the figures in the main text and SM can be accessed by \href{https://github.com/cfengno1/Moire-t-J-Model}{https://github.com/cfengno1/Moire-t-J-Model}.

\textit{Acknowledgments.---}
We thank Yuchi He for useful comments. This work  was supported by  the U.S. Department of Energy, Office of Basic Energy Sciences under Grant No. DE-FG02-06ER46305.
ITensor library~\cite{fishman_itensor_2022} is used in this work for all DMRG calculations.

\clearpage

\widetext
\begin{center}
	\textbf{\large Supplemental Materials}
\end{center}
\vspace{1mm}
\renewcommand\thefigure{S\arabic{figure}}
\renewcommand\theequation{S\arabic{equation}}
\setcounter{figure}{0} 
\setcounter{equation}{0} 
\setcounter{section}{0}

In the Supplemental Materials, we provide additional results to support the claims made in the main text. 
In Sec.~\ref{Zero_Sz}, we verify that the ground state satisfies $\sum_i \hat{S}_i^z=0$.
In Sec.~\ref{Sign_pair}, we present the relative signs for pairings along different bonds, from which one can deduce the pairing symmetries.
In Sec.~\ref{Spin_Strfac}, the spin structure factors for $\theta=\pi/12$ and $7\pi/12$ are given to show the underlying in-plane $120^{\circ}$ N\'eel and ferromagnetic orders respectively. 
In Sec.~\ref{Exp_G}, the Green's functions $G(r)$ are fitted by exponential decay. 
Sec.~\ref{Transform} shows the transformation between the STPSC and FMTSC phases. 
Sec.~\ref{sec_Ly3} shows the power-law decay of the pairing and CDW correlations on a width-3 cylinder.
Sec.~\ref{Gray_cdw} shows the power-law fittings for the pairing correlations in the gray area (non-SC regime) of the phase diagram Fig.~\ref{Fig1} in the main text, where large Luttinger exponents are found. 
Finally, in Sec.~\ref{sec_XC4} the spin structure factors and different correlation functions of the XC4 cylinder are shown, exhibiting the same phases as those in the YC4 cylinder.

\subsection{Verification of $\sum_i\hat{S}^z_i=0$ in the ground state}
\label{Zero_Sz}
To verify that the ground state satisfies $\hat{S}_\text{tot}^z\equiv\sum_i\hat{S}^z_i=0$, we run DMRG calculates for $N=12\times 4$ systemS imposing only the particle number conservation and find $\langle \hat{S}_\text{tot}^z\rangle=\langle (\hat{S}_\text{tot}^z)^2\rangle=0$ in the ground state.

\subsection{Sign structure of the pairing orders}
\label{Sign_pair}
\begin{figure}[!htbp]
\includegraphics[width=0.9\textwidth]{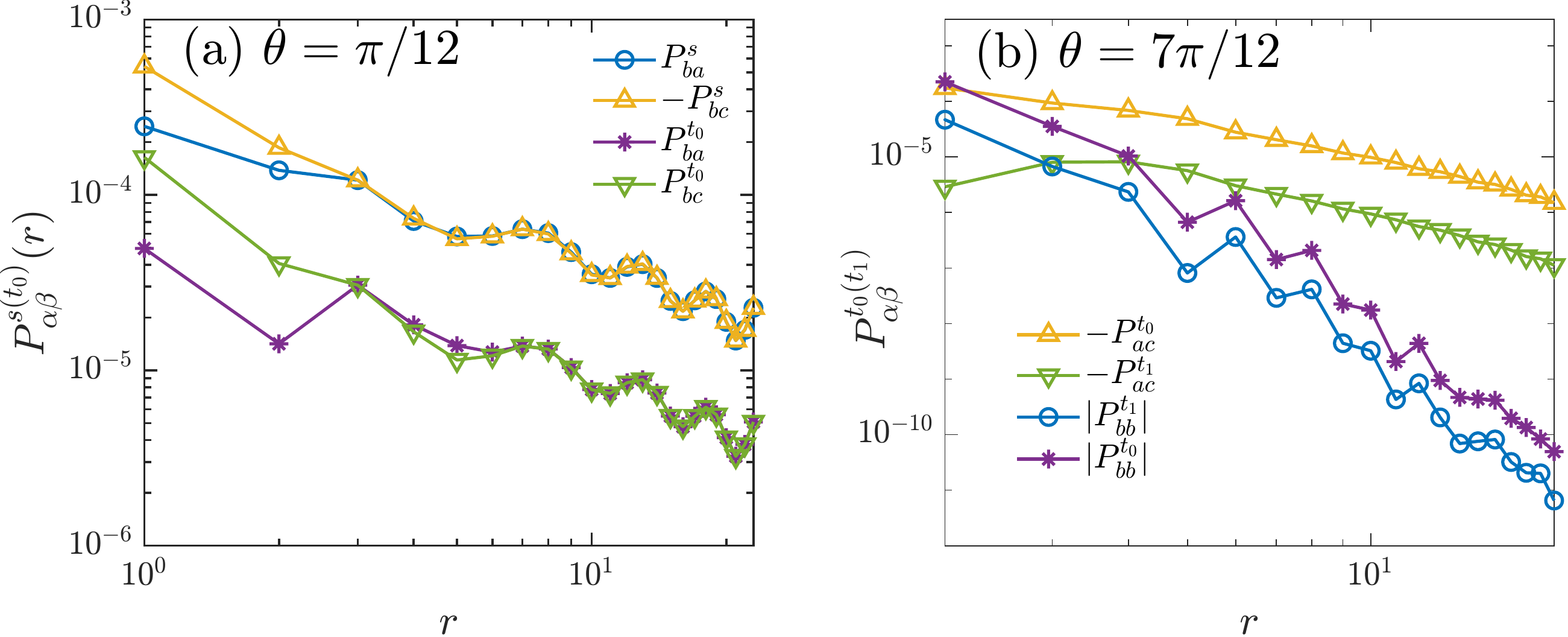}
\centering
\caption{The correlation functions between pairings along different bonds for (a) $\theta=\pi/12$ and (b) $\theta=7\pi/12$, from which one can deduce their relative sign and thus the pairing symmetry. Data are obtained under $M=17000$ and 15000 for (a) and (b) respectively in the YC4 cylinders.}
\label{Sign}
\end{figure}

We demonstrate the relative sign between pairings along two nearest-neighbor bonds $\alpha$ and $\beta$ by the sign of their correlation $P^{s(t)}_{\alpha\beta}(r)$. Specifically, for the STPSC phase represented by $\theta=\pi/12$ in Fig.~\ref{Sign}(a), we deduce sign($\Delta^s_a$)=sign($\Delta^s_b$)=$-$sign($\Delta^s_c$) and sign($\Delta^{t_0}_a$)=sign($\Delta^{t_0}_b$)=sign($\Delta^{t_0}_c$), corresponding to ordinary $d$-wave and $p$-wave symmetry respectively. For the FMTSC phase represented by $\theta=7\pi/12$ in Fig.~\ref{Sign}(b), we deduce  $\Delta^{t_0(t_1)}_b=0$ from the very weak $b$-bond triplet pairing correlations $P^{t_0(t_1)}_{bb}(r)$. Besides, we find sign($\Delta^{t_0(t_1)}_a$)=-sign($\Delta^{t_0(t_1)}_c$), therefore $p$-wave symmetry is identified.

\subsection{Spin structure factors for $\theta=\pi/12$ and $7\pi/12$ in the YC4 geometry}
\label{Spin_Strfac}
The spin structure factor $S_m(k)$ is defined as 
\begin{equation}
S_m(\boldsymbol{k})=\frac{1}{N}\sum_{i,j=1}^N e^{i\boldsymbol{k}\cdot(\boldsymbol{r}_i-\boldsymbol{r}_j)}\langle \hat{S}^x_i\hat{S}^x_j+\hat{S}^y_i\hat{S}^y_j\rangle.
\end{equation}
For the geometry in Fig.~1 of the main text, the wavevector $\boldsymbol{k}$ are quantized according to
\begin{equation}
\boldsymbol{k}\cdot L_y \boldsymbol{e}_y=2n\pi, n\in \mathbb{Z}.
\end{equation}

\begin{figure}[!htbp]
\includegraphics[width=0.35\textwidth]{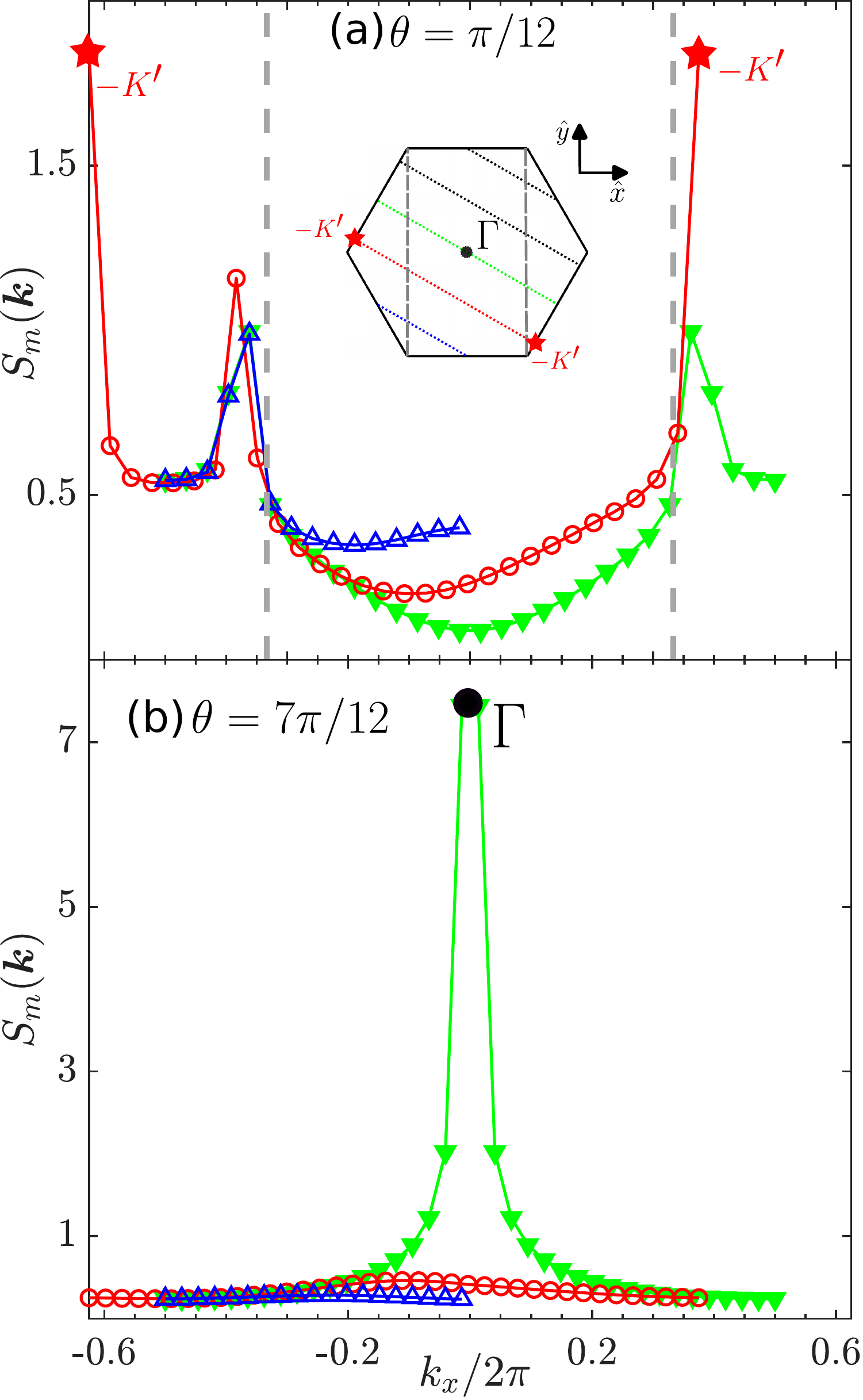}
\centering
\caption{The spin structure factors $S_m(\boldsymbol{k})$ for (a) $\theta=\pi/12$ and (b) $\theta=7\pi/12$. The inset shows the resolved wavevector $\boldsymbol{k}$s (denoted by dotted lines) in the YC4 geometry. The associated $S_m(\boldsymbol{k})$s are labeled by the respective colors. The peaks at $-\boldsymbol{K'}$ in (a) and $\boldsymbol{\Gamma}$ in (b) are inherited from the undoped system~\cite{wu_topological_2019}. The other peaks appear when x components of the wavevectors are in proximity to those of the Brillouin zone corners (i.e. $k_x\approx 2\pi/3$, marked by the gray dashed lines) and are considered as finite-size effects. Data at $M=15000$ are used.}
\label{Smk}
\end{figure}

The undoped parent state has a in-plane $120^{\circ}$ N\'eel order charaterized by peaks at Brillouin zone corners~\cite{zang_hartree-fock_2021}. Fig.~\ref{Smk}(a) shows a dominant peak at $\pm\boldsymbol{K'}$ when $\theta=\pi/12$, and since $\pm\boldsymbol{K'}$ are the nearest resolved wavevectors to the zone corners, we conclude that spin density wave order at $\theta=\pi/12$ is inherited from the parent state N\'eel order. In contrast, the structure factor is peaked at the Brillouin zone center $\boldsymbol{\Gamma}$ when $\theta=7\pi/12$,  indicating a ferromagnetic order, which is also host by the undoped system~\cite{zang_hartree-fock_2021}.

\subsection{Exponential fits for $|G(r)|$ at $\theta=\pi/12$ and $7\pi/12$ in the YC4 geometry}
\label{Exp_G}
\begin{figure}[!htbp]
\includegraphics[width=0.37\textwidth]{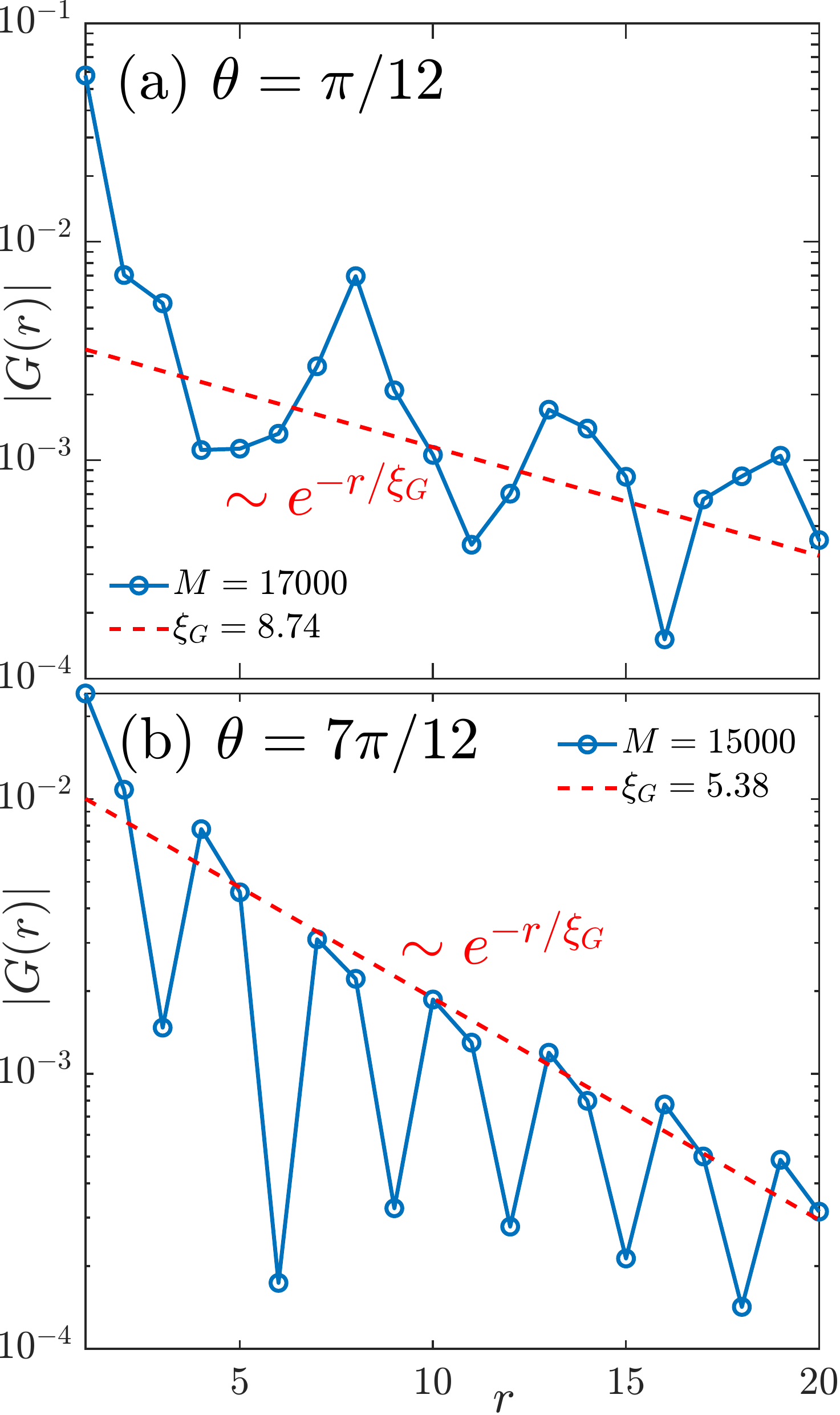}
\centering
\caption{Exponential decay of $|G(r)|$ at (a) $\theta=\pi/12$ and (b) $\theta=7\pi/12$.}
\label{Gr2}
\end{figure}

The Green's functions in the main text are fitted by power-law decays with relatively larger exponents than the pairing correlations, but here we show in Fig.~\ref{Gr2} that they can be equally well fitted by exponential decays. 

\subsection{Transformation between STPSC and FMTSC phases}
\label{Transform}
We show in this section how to transform between the FMTSC and STPSC phase by a spin-flip operation followed by a gauge transformation. First, the spin flip is tantamount to reversing $\theta$ in the Hamiltonian:
\begin{equation}
\hat{H}(\theta)\xrightarrow{\hat{\Pi}_x}\hat{H}(-\theta).
\end{equation}
Second, $\hat{H}(-\theta)$ and $\hat{H}(2\pi/3-\theta)$ are gauge equivalent as the flux through each triangle differs by $2\pi$ in them and the gauge transformation~\cite{zhou_chiral_2022} acts as:
\begin{equation}
\begin{split}
&\hat{H}(-\theta)\xrightarrow{\hat{U}_g}\hat{H}(2\pi/3-\theta),\\
&\hat{c}_\sigma(\boldsymbol{r})\xrightarrow{\hat{U}_g}e^{-i\frac{\boldsymbol{r}}{3}\cdot(\boldsymbol{b_1}-\boldsymbol{b_2})\sigma}\hat{c}_\sigma(\boldsymbol{r}),
\end{split}
\end{equation}
where $\boldsymbol{b_{1,2}}$ are the reciprocal wavevectors conjugated to $\boldsymbol{e_a}$ and $\boldsymbol{e_b}$ respectively.
Combining these two operations then gives 
\begin{equation}
\hat{H}(\theta)\xrightarrow{\hat{U}_g\hat{\Pi}_x}\hat{H}(2\pi/3-\theta).
\end{equation}
Now suppose there is uniform pairing order parameter at $\theta$: 
\begin{equation*}
\Delta^{\sigma\sigma'}_\alpha(\theta)\equiv\bra{\Psi(\theta)}\hat{c}_\sigma(\boldsymbol{r})\hat{c}_{\sigma'}(\boldsymbol{r+e_\alpha})\ket{\Psi(\theta)}\neq 0,
\end{equation*}
where $\ket{\Psi(\theta)}$ is the ground state for $\hat{H}(\theta)$. Since $\ket{\Psi(2\pi/3-\theta)}=\hat{U}_g\hat{\Pi}_x \ket{\Psi(\theta)}$, we have for $2\pi/3-\theta$:
\begin{equation}
\begin{split}
&\Delta^{-\sigma,-\sigma'}_\alpha(2\pi/3-\theta,\boldsymbol{r})\\
\equiv&\bra{\Psi(2\pi/3-\theta)}\hat{c}_{-\sigma}(\boldsymbol{r})\hat{c}_{-\sigma'}(\boldsymbol{r+e_\alpha})\ket{\Psi(2\pi/3-\theta)}\\
=&e^{-i\frac{\sigma+\sigma'}{3}(\boldsymbol{b_1-b_2})\cdot\boldsymbol{r}-i\frac{\sigma'}{3}\boldsymbol{e_\alpha\cdot(b_1-b_2)}}\Delta^{\sigma\sigma'}_\alpha(\theta),
\end{split}
\label{TF_SM}
\end{equation}
which gives rise to Eq.~5 and 6 in the main text.

\subsection{SC and CDW orders for $L_y=3$}
\label{sec_Ly3}
\begin{figure}[!htbp]
\includegraphics[width=0.35\textwidth]{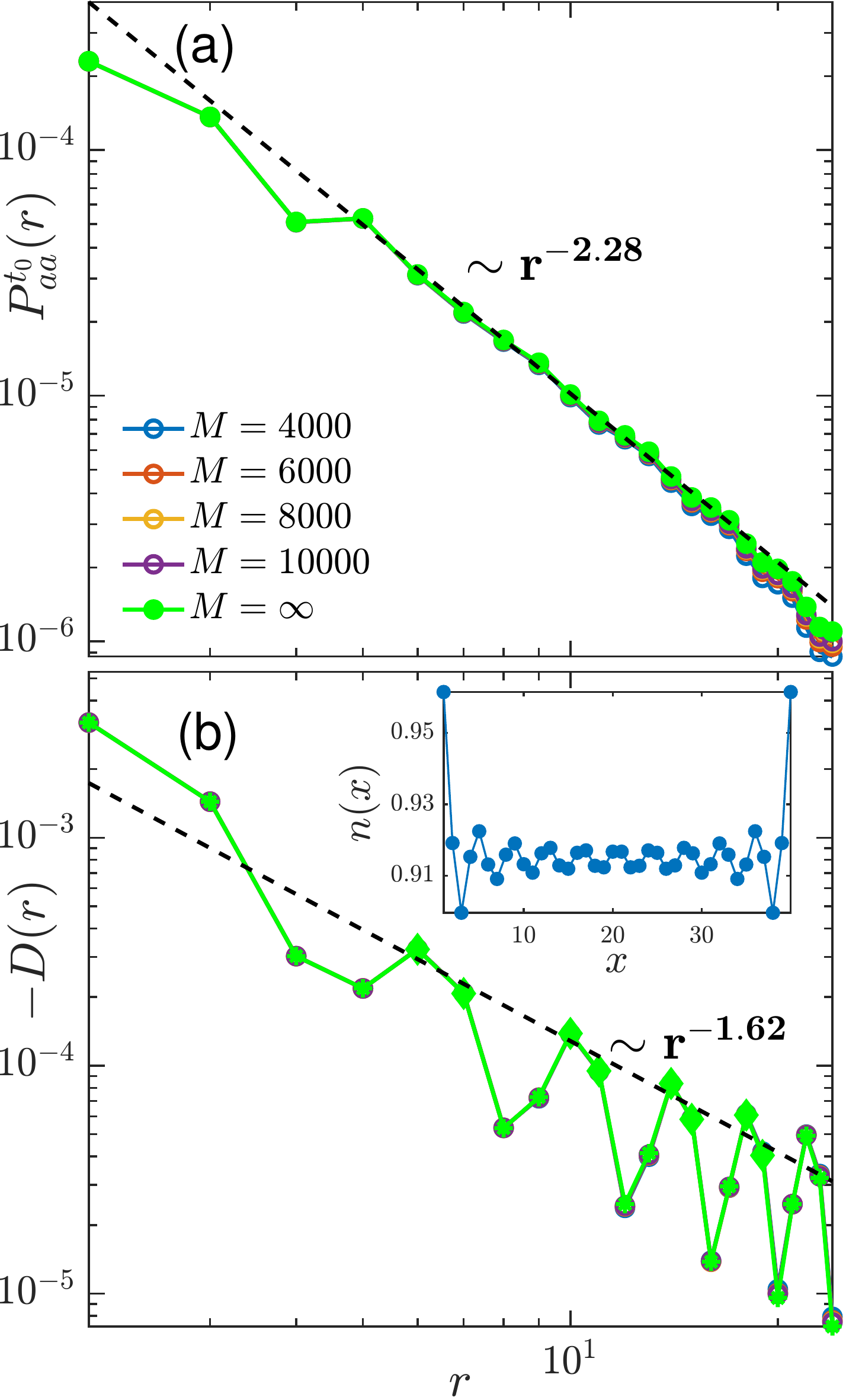}
\centering
\caption{The dominant pairing and charge density correlations $P^{t_0}_{aa}$ and $D(r)$ for a YC3 system with $N=40\times 3$, $\theta=7\pi/12$ and $\delta=1/12$. The Luttinger exponents ($K^{t_0}_\text{sc}\approx 2.28$ and $K_\text{cdw}\approx 1.62$) are very similar to those of the 4x36 system ($K^{t_0}_\text{sc}\approx 2.16$ and $K_\text{cdw}\approx 1.64$).}
\label{N3x40}
\end{figure}

To study the width dependence of the SC and CDW order, we also calculate a YC3 system at $\theta=7\pi/12,\delta=1/12$ (see Fig.~\ref{N3x40}) and find that the Luttinger exponents for pairing and charge density correlations are close to those of the YC4 system. Similarly, a charge stripe order with one hole per stripe is observed.

\subsection{SC correlation of non-SC (gray-dot) regime in the phase diagram}
\label{Gray_cdw}
\begin{figure}[!htbp]
\includegraphics[width=0.35\textwidth]{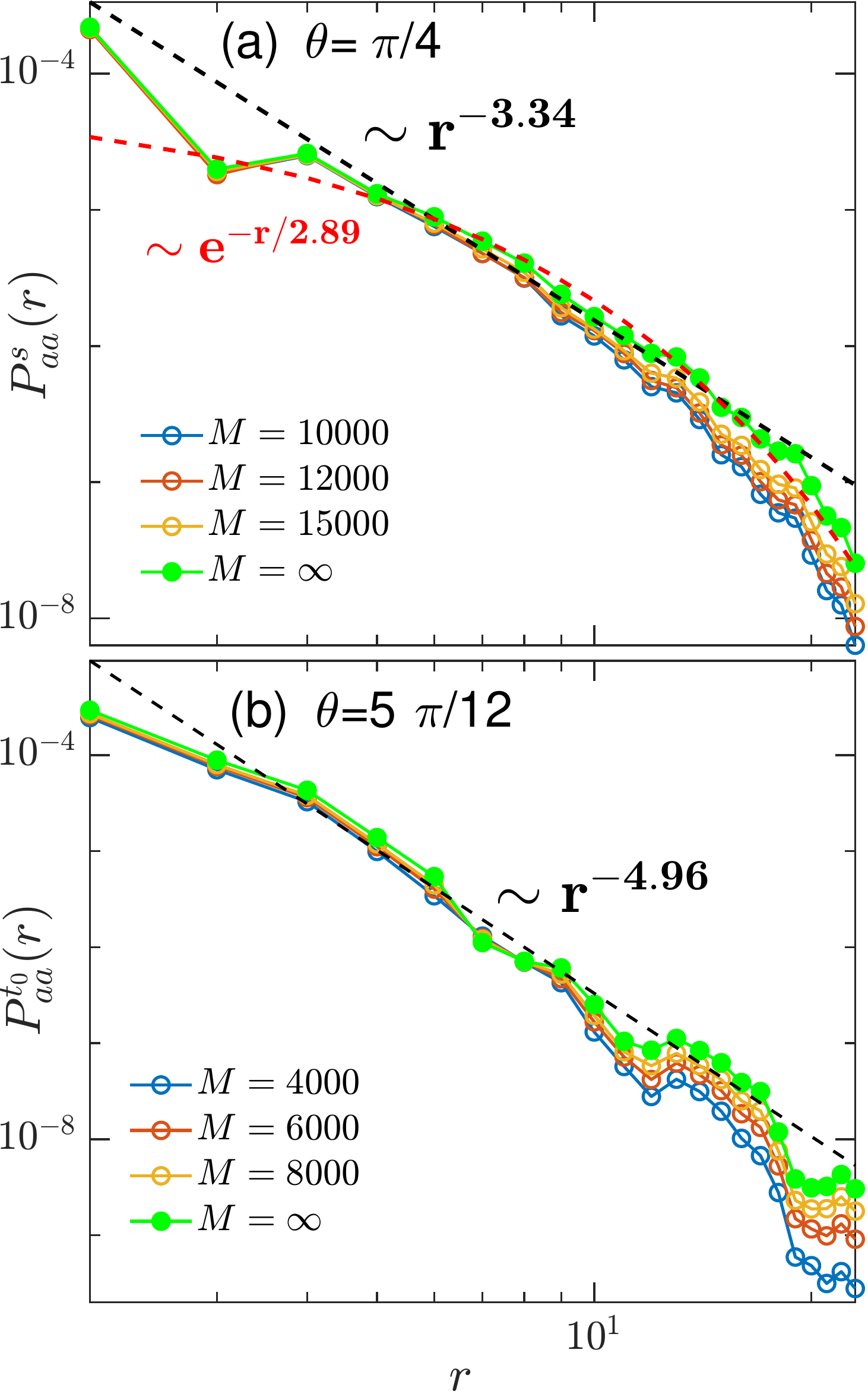}
\centering
\caption{The dominant pairing correlations for (a) $\theta=\pi/4$ and (b) $\theta-5\pi/12$. Luttinger exponents are larger than 3 are observed, and the SC correlation length ($\sim 2.89$) is smaller than the system width.}
\label{Gray_dots}
\end{figure}

We show in Fig.~\ref{Gray_dots} the dominant SC correlations for two representative points in the middle regime of the phase diagram Fig.~1(b) in the main text. The Luttinger exponents are significantly larger than 2 and the exponential fit gives a SC correlation length smaller than the system width. Therefore a clear signature of SC order is lacking.

\subsection{Results for the XC4 geometry}
\label{sec_XC4}
\begin{figure}[!htbp]
\includegraphics[width=0.42\textwidth]{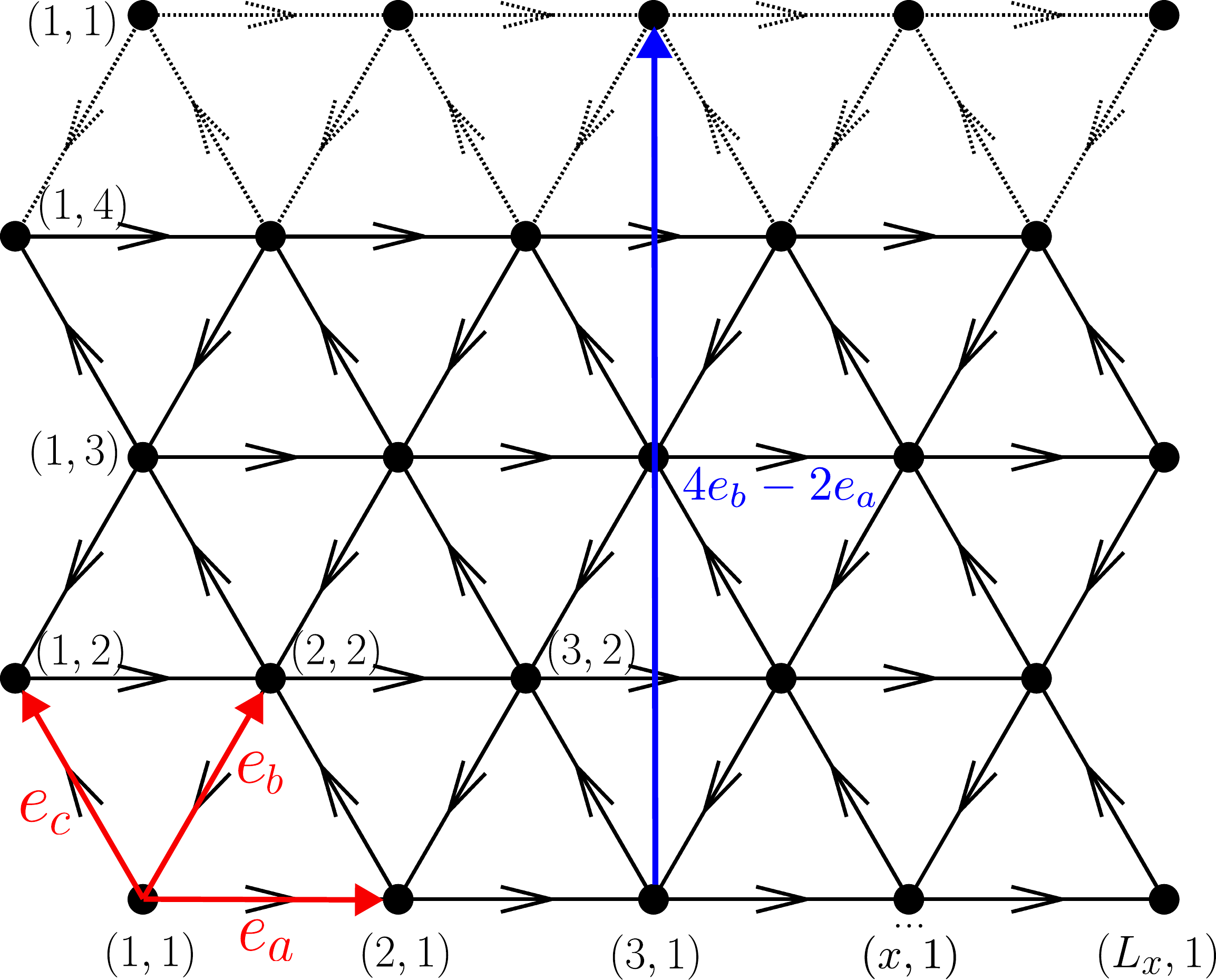}
\centering
\caption{The XC4 geometry. The first and last row are identified together. The system is periodic under the translation $4\boldsymbol{e_b}+2\boldsymbol{e_a}$. The $x$-th site counted from the left edge on the $y$-th row is denoted by $(x,y)$. }
\label{XC4}
\end{figure}

\begin{figure}[!htbp]
\includegraphics[width=0.45\textwidth]{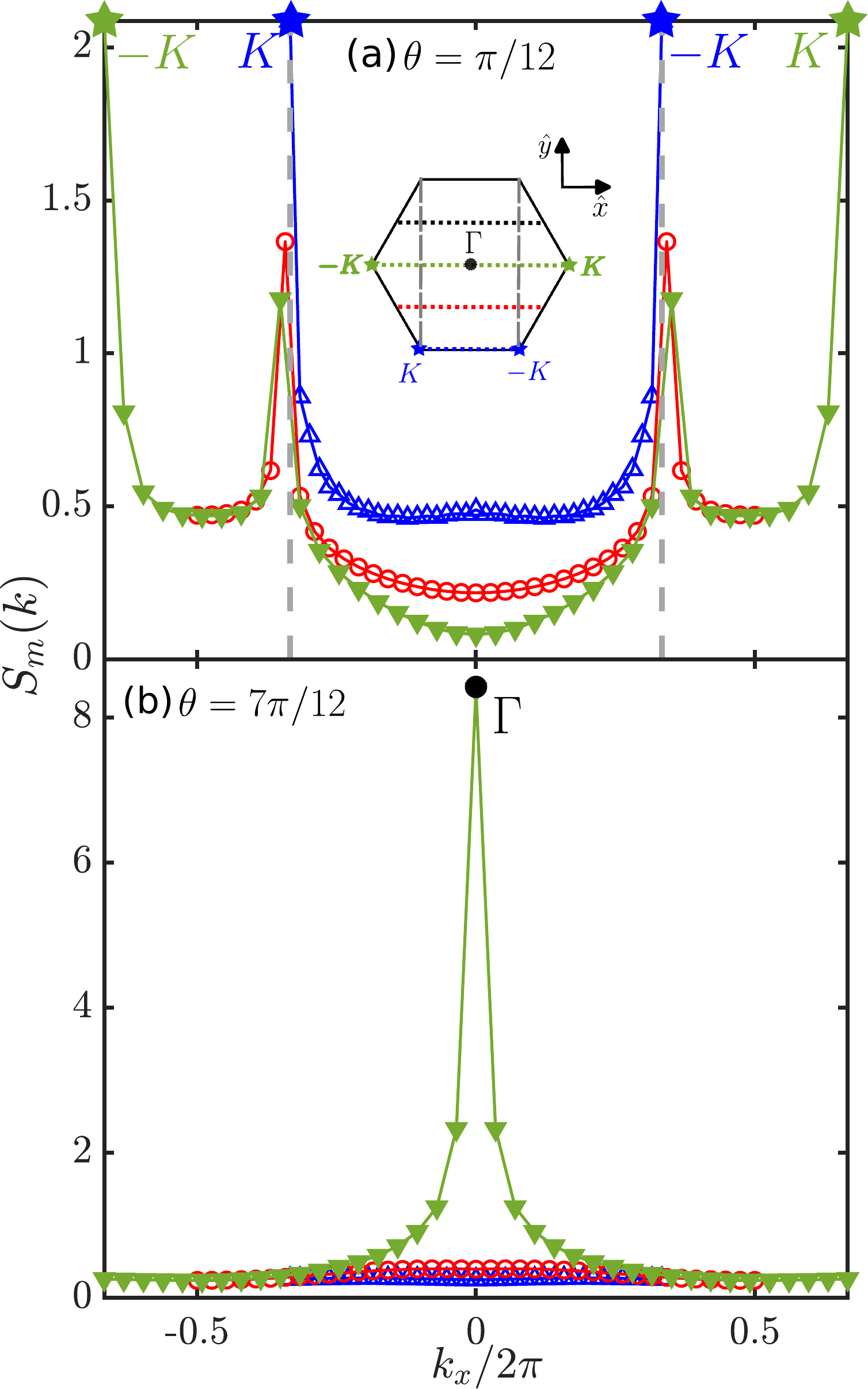}
\centering
\caption{The spin structure factors $S_m(\boldsymbol{k})$ for (a) $\theta=\pi/12$ and (b) $\theta=7\pi/12$ in the XC4 geometry. The inset shows the resolved wavevector $\boldsymbol{k}$s denoted by the dotted lines in the Brillouin zone. The associated $S_m(\boldsymbol{k})$s are labeled by the respective colors. The peaks at $\pm\boldsymbol{K}$ in (a) and $\boldsymbol{\Gamma}$ in (b) are inherited from the undoped system~\cite{wu_topological_2019}. The other peaks appear when x components of the wavevectors are in proximity to those of the Brillouin zone corners (i.e. $k_x\approx 2\pi/3$, marked by the gray dashed lines), and are considered as finite-size effects. Data at $M=12000$ are used.}
\label{Smk_XC}
\end{figure}

To complement the YC4 geometry in the main text, here we also study another geometry called XC4~\cite{szasz_chiral_2020} shown in Fig.~\ref{XC4}. The wavevector $\boldsymbol{k}$s are quantized according to
\begin{equation}
\boldsymbol{k}\cdot(4\boldsymbol{e_b}-2\boldsymbol{e_a})=2n\pi, n\in\mathbb{Z},
\end{equation}
and the resolved $\boldsymbol{k}$s are denoted as dotted line in the inset of Fig.~\ref{Smk_XC}(a). The PBC 
\begin{equation}
\hat{c}_{\boldsymbol{r}+4\boldsymbol{e_b}-2\boldsymbol{e_a},\sigma}=\hat{c}_{\boldsymbol{r}\sigma}
\end{equation}
is unchanged under gauge transformation.

The spin structure factors for $\theta=\pi/12$ and $7\pi/12$ in Fig.~\ref{Smk_XC} shows main peaks at $\pm\boldsymbol{K}$ and $\boldsymbol{\Gamma}$ respectively, confirming the $120^{\circ}$ N\'eel and ferromagnetic orders inherited from the parent states.

\begin{figure}[!htbp]
\includegraphics[width=0.6\textwidth]{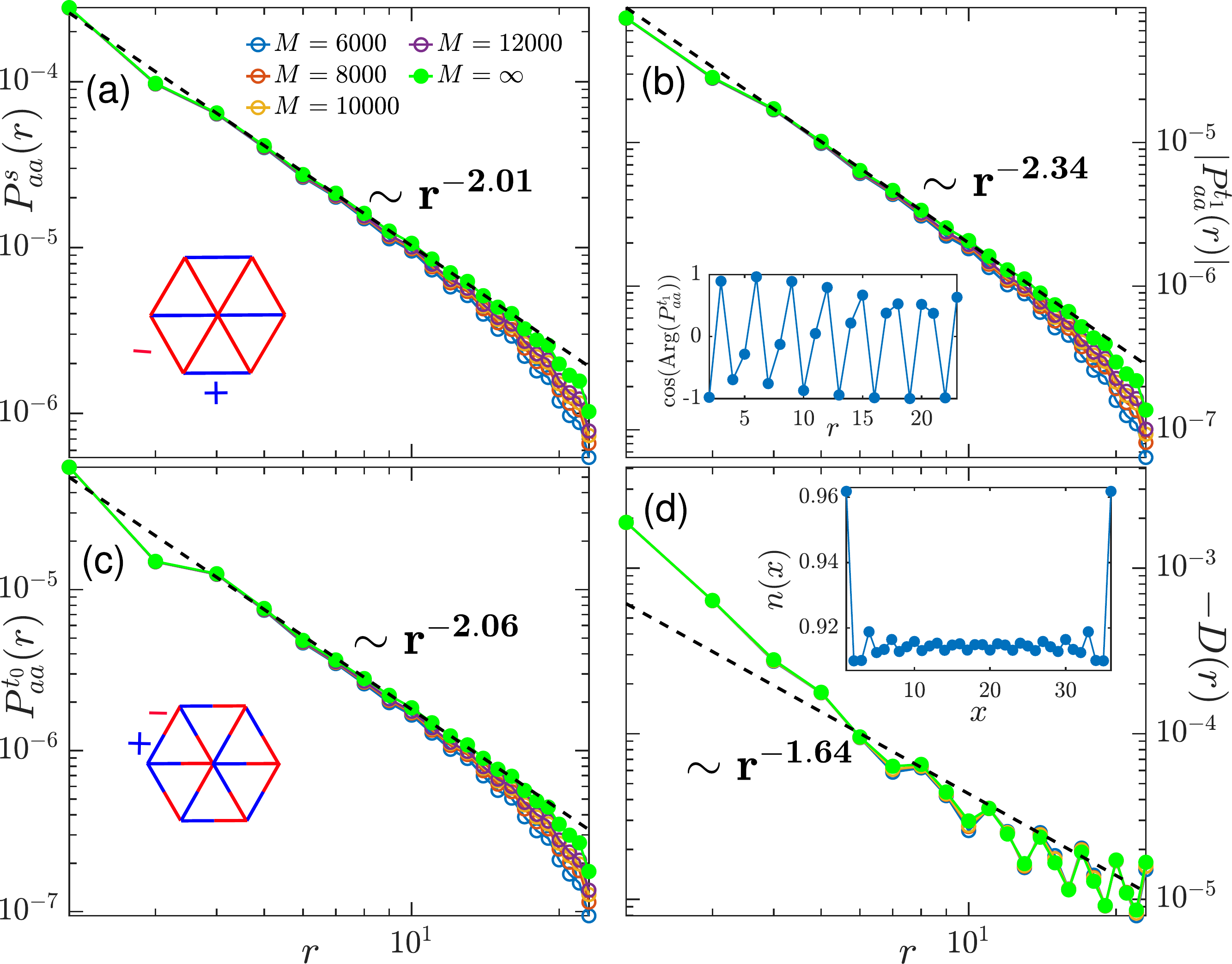}
\centering
\caption{STPSC phase at $\theta=\pi/12$ in the XC4 geometry. Scaling and fitting of the singlet pairing correlation (a), triplet paring correlation in the same-spin (b) and opposite-spin channel (c). $d$-wave symmetry of the singlet pairing and $p$-wave symmetry of the triplet pairing are identified by the sign structures in the insets of (a) and (c). The inset of (b) shows the cosine of the argument of the complex-valued same-spin triplet pairing correlation, whose spatial oscillation reveals the PDW order. (d) The density-density correlation. The inset shows the electron density profile $n(x)=\sum_y\langle \hat{n}(x,y)\rangle/L_y$ and a weak charge stripe order with average one hole per stripe is observed. The bond dimension is kept up to $M=12000$, corresponding to the truncation error around $1\times 10^{-7}$.}
\label{Pi_12_XC}
\end{figure}

\begin{figure}[!htbp]
\includegraphics[width=0.6\textwidth]{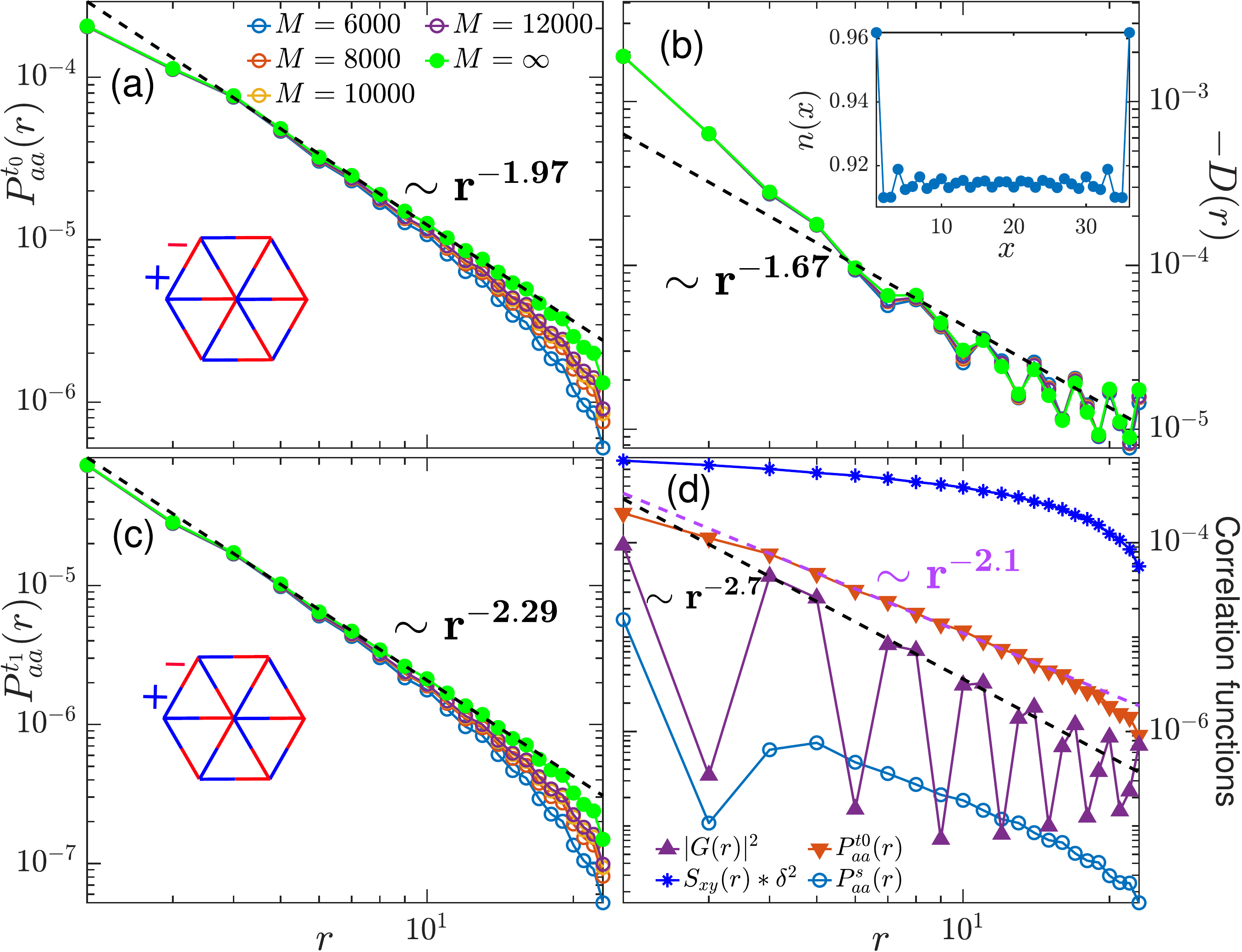}
\centering
\caption{FMTSC phase at $\theta=7\pi/12$ in the XC4 geometry. Scaling and fitting of the triplet paring correlation in the opposite-spin (b) and same-spin channel (c) $p$-wave symmetry of both triplet pairing channels are determined by the sign structures in the insets. (b) The density-density correlation and electron density profile, which are identical with the those in Fig.~\ref{Pi_12_XC}(b) . (d) Comparison between different correlations. The ferromagnetic spin correlation is the dominant order. The singlet pairing is negligible compared to the triplet one. The triplet pairing correlation decays slower than the Green's function square, suggesting the dominance of two-electron over single electron transport process.  The bond dimension is kept up to $M=12000$, corresponding to the truncation error around $1\times 10^{-7}$.}
\label{7Pi_12_XC}
\end{figure}

Similar to the YC4 geometry, we also find the STPSC phase at $\theta=\pi/12$ (Fig.~\ref{Pi_12_XC}) and FMTSC phase at $\theta=7\pi/12$ (Fig.~\ref{7Pi_12_XC}) in the XC4 geometry with SC Luttinger exponents around 2. Note that the Luttinger exponents satisfy $K^s_\text{sc}(\theta=\pi/12)\approx K^{t_0}_\text{sc}(\theta=\pi/12)\approx K^{t_0}_\text{sc}(\theta=7\pi/12)$. This is expected due to the linear relation in Eq.~5 in the main text and the near absence of singlet pairing at $\theta=7\pi/12$ in Fig.~\ref{7Pi_12_XC}(d). Likewise $K^{t_1}_\text{sc}(\theta=\pi/12)\approx K^{t_1}_\text{sc}(\theta=7\pi/12)$, in agreement with the relation in Eq.~\ref{TF_SM}.

\end{document}